\documentstyle[12pt]{article}

\newcommand{\EQ}{\begin{equation}}
\newcommand{\EN}{\end{equation}}

\newcommand{\bear}{\begin{eqnarray}}
\newcommand{\ear}{\end{eqnarray}}

\begin{document}

\topmargin 0pt
\oddsidemargin 5mm
\newcommand{\NP}[1]{Nucl.\ Phys.\ {\bf #1}}
\newcommand{\PL}[1]{Phys.\ Lett.\ {\bf #1}}
\newcommand{\NC}[1]{Nuovo Cimento {\bf #1}}
\newcommand{\CMP}[1]{Comm.\ Math.\ Phys.\ {\bf #1}}
\newcommand{\PR}[1]{Phys.\ Rev.\ {\bf #1}}
\newcommand{\PRL}[1]{Phys.\ Rev.\ Lett.\ {\bf #1}}
\newcommand{\MPL}[1]{Mod.\ Phys.\ Lett.\ {\bf #1}}
\newcommand{\JETP}[1]{Sov.\ Phys.\ JETP {\bf #1}}
\newcommand{\TMP}[1]{Teor.\ Mat.\ Fiz.\ {\bf #1}}
     
\renewcommand{\thefootnote}{\fnsymbol{footnote}}
     
\newpage
\setcounter{page}{0}
\begin{titlepage}     
\begin{flushright}
UFSCAR-TH-00-25
\end{flushright}
\vspace{0.5cm}
\begin{center}
\large{ The excitations of the symplectic integrable models and their applications} \\
\vspace{1cm}
\vspace{1cm}
 {\large M.J. Martins } \\
\vspace{1cm}
\centerline{\em Departamento de F\'isica, 
Universidade Federal de S\~ao Carlos}
\centerline{\em Caixa Postal 676, 13565-905, S\~ao Carlos, Brasil}
\vspace{1.2cm}   
\end{center} 
\begin{abstract}
The Bethe ansatz equations of the fundamental $Sp(2N)$ integrable
model are solved by a peculiar configuration of roots 
leading us to determine the nature of the excitations. 
They consist of $N$ elementary generalized spinons and $N-1$ composite
excitations made by special convolutions between the spinons.
This fact is essential
to determine the low-energy behaviour which is argued to be described
in terms of $2N$ Majorana fermions. Our results have practical 
applications to spin-orbital systems and also shed new light
to the connection between integrable models and Wess-Zumino-Witten field
theories.
\end{abstract}
\vspace{.2cm}
\centerline{PACS numbers: 05.50.+q, 75.10.Jm }
\vspace{.2cm}
\centerline{December 2000}
\end{titlepage}

\renewcommand{\thefootnote}{\arabic{footnote}}

The study of quantum one-dimensional integrable models has turned out
to be a fruitful venture since the seminal work of Bethe in 1931 \cite{BE}.
Over the years, solvable models have been extremely useful in many
subfields of physics, providing us a rich laboratory in which new 
theoretical insights and non-perturbative methods can readily be tested.
Recent progress in the experimental study of low-dimensional materials,
e.g. spin ladders and carbon nanotubes \cite{LA}, has been an additional
source of motivation to investigate one-dimensional exactly solvable
models.

The basic concept of quantum integrability is the $S$-matrix which represents
either the factorized scattering of particles of (1+1) quantum field
theories or the statistical weights of integrable
two-dimensional lattice models. It turns out that the symmetry of the
$S$-matrix plays a fundamental role in the theory and classification
of integrable systems \cite{SP1,BER}. Of particular interest is the
$Sp(2N)$ symmetry which preserves bilinear antisymmetric metrics,
typical of systems with $N$ component Dirac fermions. Even though the
Bethe ansatz solution of the integrable $Sp(2N)$ models has long
been known \cite{SP2,MA}, basic properties such as the nature
of the elementary excitations and the low-energy behaviour have not
yet been determined. 

The purpose of this paper is to unveil
the physical content of the  fundamental (vector representation)
$Sp(2N)$ solvable magnet.
We argue that the low-energy properties
are given in terms of $2N$ Majorana fermions due to the presence of
special
low-lying excitations in the spectrum. There exists at least two
immediate applications of this result. First, it provides us a
unique counterexample to the conjecture that integrable models
based on the vector representation of Lie algebras should
be lattice realizations of Wess-Zumino-Witten (WZW) conformal
theories \cite{LW1,LW2,LW3}. In fact, we predict $c=N$ for the
fundamental integrable $Sp(2N)$ model while the central charge
of the $Sp(2N)$ WZW theory is $c=N(2N+1)/(N+2)$. Next, our
study is of utility to one-dimensional systems with coupled
spin and orbital degrees of freedom such as the spin-orbital \cite{KU,SO,SO1}
and spin-tube \cite{AN} models. More precisely, we recall that the effective 
spin-isospin Hamiltonian describing these systems may be written 
in the form \cite{KU,AR}
\begin{equation}
H_{SO}(J_0,J_1,J_2) = \sum_{i=1}^{L} \sum_{\alpha=0}^{2} J_{\alpha} P^{(\alpha)}_{i,i+1}
\end{equation}
where $J_{\alpha}$ are superexchange constants and $P^{(\alpha)}_{i,i+1}$
denote the respective projections on the singlet, triplet and doublet
spin-isospin states, see ref.\cite{AR} for details. 
By writing these projectors in terms of two commuting sets of
Pauli matrices it is not difficult to identify that the
integrable $Sp(4)$ spin chain \cite{BA} corresponds to the point
$J_{0}/J_1=J_0/J_2=1/3$.
This point is interesting because it
corresponds to both anisotropic and asymmetric spin-1/2 couplings \cite{REM},
being closer to represent the properties of
realistic materials \cite{ISO,MI} than
the integrable $SU(4)$ case $J_0=J_1=J_2$ \cite{SU}. This then
provides us a rare opportunity to determine exactly the
nature of the excitations in a relevant spin-orbital
model. 

In the context of statistical mechanics the integrable $Sp(2N)$ model
is a multistate vertex system defined on the square lattice whose
bonds variables take $2N$ possible values. For each type of
configuration of four bonds $a,b,c,d$ meeting at a vertex
we associate a Boltzmann weight factor $S_{ab}^{cd}(\lambda)$ where
$\lambda$ is the spectral parameter. Compatibility between integrability
and the $Sp(2N)$ invariance (``ice-type'' restriction) leads us 
to the following amplitudes \cite{BER,MA}
\begin{equation}
S_{ab}^{cd}(\lambda)= \delta_{a,d} \delta_{c,b}
+\lambda \delta_{a,c} \delta_{b,d} -\frac{\lambda}{\lambda+N+1} \epsilon_a
\epsilon_c \delta_{a,\bar{b}} \delta_{c,\bar{d}}
\end{equation}
where ${\bar{a}}=2N+1-a$, $\epsilon_a=1$ for $1 \leq a \leq N$ and $\epsilon_a=-1$ for
$N+ 1 \leq a \leq 2N$. 

For any integrable vertex model one can associate a local spin chain
commuting with the corresponding transfer-matrix whose matrix elements
are given by ordered product of $L$ factors $S_{ab}^{cd}(\lambda)$.
As usual, the Hamiltonian is proportional to the logarithmic derivative
of the transfer matrix at the regular point $\lambda=0$, and in this case the
expression is
\begin{equation}
H_{Sp(2N)}= \sum_{i=1}^{L} \left [ 
\delta_{a,d} \delta_{c,b}
-\frac{1}{N+1} \epsilon_a
\epsilon_c \delta_{a,\bar{b}} \delta_{c,\bar{d}} \right ] e^{(i)}_{ac} \otimes
e^{(i+1)}_{bd} 
\end{equation}
where $\epsilon_{ab}^{(i)}$ are the elementary matrix 
$[e_{ab}]_{l,k}=\delta_{a,l} \delta_{b,k}$ acting
on site $i$. We observe that the
spectrum of $H_{SO}(J_1/3,J_1,J_1)$ matches that of $J_1 H_{Sp(4)} -J_1 L$.
The $Sp(2N)$ Hamiltonian (3) is solvable by the Bethe ansatz \cite{SP2,MA}
and its eigenvalues $E(L)$ can be parametrized in terms of a set of variables
$\lambda_j^{(a)}$, 
$j=1, \cdots, m_{a}$ and $a=1, \cdots, N$,
satisfying the following Bethe
equations
\EQ
\left[
\frac{\lambda^{(a)}_{j} -i\frac{\delta_{a,1}}{\eta_{a}}}
{\lambda^{(a)}_{j} +i\frac{\delta_{a,1}}{\eta_{a}}} 
\right]^{L} =
\prod_{b=1}^{N} \prod_{k=1,\; k \neq j}^{m_{b}}
\frac{\lambda^{(a)}_{j}-\lambda^{(b)}_{k} -i\frac{C_{a,b}}{\eta_{a}}}
{\lambda^{(a)}_{j}-\lambda^{(b)}_{k} +i\frac{C_{a,b}}{\eta_{a}}}, ~~ 
\EN
and the eigenvalues are given by 
\EQ
E(L) = - \sum_{i=1}^{m_{1}} \frac{1}{\left [ \lambda^{(1)}_{i}\right ]^{2} + 1/4} + L
\EN
where $C_{ab}$ is the Cartan matrix 
and $\eta_a$ is the normalized length of the
$a$-th root of the $Sp(2N)$ algebra. 

We start our study considering first the $Sp(4)$ model motivated
by its direct relevance to the physics of spin-orbital systems. In fact,
this is the simplest symplectic invariant system since $N=1$ is 
equivalent
to the isotropic six-vertex model. Later on  we will
show that the technicalities entering in the
analysis of the $Sp(4)$ model can be easily generalized to include arbitrary
$N>2$. Essential to our study is to determine the configurations
of roots that describe the absolute ground state and the elementary
excitations. This can be done by solving the Bethe equations (4,5) for
some values of $L$ and comparing it with the exact diagonalization of
the $Sp(4)$ spin chain. Its spectrum is parametrized by two set
of variables $\lambda_j^{(1,2)}$ and we found that the ground state
and the low-lying excitations are characterized by strings with
different lengths, namely
\begin{equation}
\lambda_{j}^{(1)} = 
\left\{\begin{array}{l}
\xi_j^{(1)} \\
\xi_j^{(3)} \pm i \left [\frac{1}{2} +O(e^{-\gamma L}) \right ]
\end{array}\right. ~~~ \lambda_j^{(2)}=\xi_j^{(2)}
\end{equation}
with $\gamma$ positive and $\xi_j^{(\alpha)}$ $\alpha=1,2,3$ are real numbers. 

The important point here is to observe that the rapidities $\lambda_j^{(1)}$
are described in terms of two independent type of strings, i.e 1-strings
and 2-strings. This feature should be contrasted to the behaviour of the Bethe
ansatz roots of other fundamental integrable systems such as $SU(N)$
and $O(N)$ models \cite{SU,LW2}. In fact, for the latter systems each
$a$-th root $\lambda_j^{(a)}$ characterizing the infinite volume
properties has 
a $unique$ string length. To explore the consequences of the peculiar
string configuration (6) we substitute it in the Bethe ansatz (4) and
we obtain the following effective equations
for the variables $\xi_{j}^{\alpha}$
\begin{equation}
L \left [ \psi_{1/2}(\xi_j^{(\alpha)}) \delta_{\alpha,1}+
\psi_{1}(\xi_j^{(\alpha)}) \delta_{\alpha,3} \right ]= 2 \pi Q_j^{(\alpha)} +
\sum_{\beta=1}^{3} \sum_{k=1}^{N_{\beta}} \phi_{\alpha \beta}( 
\xi_j^{(\alpha)}-\xi_j^{(\beta)})
\end{equation}
where $\psi_{a}(x)=2 \arctan(x/a)$, $N_{\alpha}$ is the number of roots
$\xi_j^{(\alpha)}$, $Q_j^{(\alpha)}= -(N_{\alpha}-1)/2 +j-1$ with $j=1, \cdots, 
N_{\alpha}$ and the matrix elements 
$\phi_{\alpha \beta}(x)$ are given by 
\begin{equation}
\phi_{\alpha \beta}(x)=\left(\matrix{\psi_1(x )& -\psi_1(x)
&\psi_{1/2}(x)+\psi_{3/2}(x)\cr
-\psi_1(x)&\psi_2(x)&-\psi_{1/2}(x)-\psi_{3/2}(x)\cr 
\psi_{1/2}(x)+\psi_{3/2}(x)&-\psi_{1/2}(x)-\psi_{3/2}(x)
&2\psi_1(x) +\psi_2(x)\cr }\right)
\end{equation}

The ground state for even $L$ consists of a sea of $1$-strings and $2$-strings
with $N_1=N_2=2N_3=L/2$. For large $L$, the roots $\xi_j^{(a)}$
are densely packed into its density distribution 
$\sigma(\xi_j^{(\alpha)})=1/L(\xi_{j+1}^{(\alpha)}-\xi_j^{(\alpha)})$ 
and the relations (7)
in the $L \rightarrow \infty$ limit become integral 
equations for such densities. These integral equations 
are solved by elementary Fourier techniques and we find that
\begin{equation}
\sigma^{(1)}(x) =\frac{1}{2 \cosh(\pi x)},~~ 
\sigma^{(2)}(x) =\frac{1}{6 \cosh(\pi x/3)} 
\end{equation}
and
\begin{eqnarray}
\sigma^{(3)}(x) &= &\frac{1}{2 \pi} \int_{-\infty}^{+\infty} 
\sigma^{(1)}(y) \sigma^{(2)}(x-y)dy  \nonumber \\
& &
=\frac{ 
\frac{2}{\sqrt{3}} \sinh(2 \pi x/3) -x}{6 \sinh(\pi x)}
\end{eqnarray}
where we have emphasized the remarkable fact that $\sigma^{(3)}(x)$ is
exactly the convolution of the densities $\sigma^{(1)}(x)$ and
$\sigma^{(2)}(x)$. 
Recall that function
$\sigma^{(\alpha)}(x)$ is related to the continuous probability 
densities of finding the rapidity $\xi^{(\alpha)}$ with a given value $x$.
We may therefore interpret $\sigma^{(3)}(x)$ as the probability
for the sum of two independent events with probability $\sigma^{(1)}(y)$
at an arbitrary value $y$ and $\sigma^{(2)}(x-y)$ at the
complementary value $x-y$. 

We now have the basic ingredients to investigate the thermodynamic
limit properties. The ground state 
energy per site $e_{\infty}$ is calculated
by using Eqs.(5,9,10) after replacing the sum by an integral. The 
final result is
\begin{equation}
e_{\infty}
=-2\left [ \frac{2 \ln(2)}{3} 
+\frac{\pi}{9 \sqrt{3}} -\frac{1}{3} \right ] 
\end{equation}

The low-lying excitations are obtained by inserting holes in the density
distribution of $\xi_j^{(\alpha)}$ which means the removal of certain
quantum numbers $Q_j^{(\alpha)}$ of the Bethe equations (7). The necessary
manipulations of these equations are standard \cite{FA,SU}
and we find that the energy
$\varepsilon^{(\alpha)}(\xi)$ and the momentum $p^{(\alpha)}(\xi)$ 
of one hole excitation in the sea of $\xi_j^{(\alpha)}$, measured
from the ground state, have the form
\begin{equation}
\varepsilon^{(\alpha)}(\xi)= \pi \sigma^{(\alpha)}(\xi), ~~~
p^{(\alpha)}(\xi) = \int_{\xi}^{+\infty} \varepsilon^{(\alpha)}(x) dx
\end{equation}
where $\xi$ is the $\alpha$-hole rapidity. For the first two excitations one can
easily eliminate the variable $\xi$ leading us to the following
dispersion relations
\begin{equation}
\varepsilon^{(1)}(p)= \frac{\pi}{2} \sin(2p), ~~~
\varepsilon^{(2)}(p)= \frac{\pi}{6} \sin(2p)
\end{equation}
implying that these excitations are gapless and that their low-energy
limits $\varepsilon^{(\alpha)}(p) \sim v^{(\alpha)} p $ are governed by
distinct sound velocities, i.e $v^{(1)}=\pi$ and $v^{(2)}=\pi/3$. The
contribution to the total spin of each of these excitations is
$\frac{1}{2}$, and therefore we shall interpret them as
spinons propagating with different velocities.

Similar computation for the third excitation leads us to transcendental
equations and an analytical expression for the dispersion relation is hard
to be obtained. 
However, it is possible to study the low-energy behaviour of
such excitation by expanding the density $\sigma^{(3)}(\lambda)$ in 
powers of $e^{-\lambda}$. We see that low-momenta regime is dominated
by both sound velocities $v^{(1)} $ and $v^{(2)}$ and strictly 
in the $p \rightarrow 0$ limit the lowest one prevails. This massless
excitation turns out to be a spinless mode whose speed of sound is
$v^{(3)}=\pi/3$.
At this point we  
note that recently the compound $NaV_2O_5$ has been modeled
by an anisotropic/asymmetric spin-orbital model \cite{ISO}. Remarkably,
the three-particle continuum found above is in accordance with
the excitation spectrum proposed in ref.\cite{ISO} to explain the optical
properties of this material.

Next we would like to identify the underlying conformal field theory which
describes the low-energy limit of the integrable $Sp(4)$ model. This
can be investigated by analyzing the behaviour of the finite size
spectrum of the corresponding spin chain. For a conformally invariant
theory,
the ground state energy $E_0(L)$ in a finite lattice of
size $L$ behaves as \cite{CA,KO}
\begin{equation}
\frac{E_0(L)}{L}= e_{\infty} -\frac{\pi}{6 L^2}\sum_{a=1}^{3} v^{(a)}c^{(a)} 
\end{equation}
where $c^{(a)}$ is the central charge associated to the $a$-th excitation.
Similarly, the conformal dimension $X_{i}^{(a)}$ of the operator
corresponding to the excited state $E_i(L)$ is proportional to the finite size
gap
\begin{equation}
\frac{E_i(L)}{L} -\frac{E_0(L)}{L}= \frac{2\pi}{ L^2} \sum_{a=1}^{3}
v^{(a)}X^{(a)}_i
\end{equation}

Within the string hypothesis (6), the finite size corrections to the
eigenspectrum can be evaluated analytically by applying the root density
method \cite{WA} to the ``string'' Bethe equations (7). For instance, we
find that each $a$-th massless excitation is associated with a central
charge $c^{(a)}=1$. However, the string assumptions gives us only
an idea of the true finite size behaviour, because the complex part
of the roots may contribute to the term $1/L^2$ as well.  In order
to obtain the correct finite size properties, we solve numerically
the original Bethe ansatz equations (4) up to $L=36$. In table (1) 
we exhibit such numerical results for the effective central charge
$c_{ef}=\sum_{i=1}^{3}\frac{v^{(i)}}{v^{(2)}}c^{(i)}$ and the
conformal dimension 
$X_{ef}=\sum_{i=1}^{3}\frac{v^{(i)}}{v^{(2)}} X^{(i)}_{1}$ 
associated
with the lowest spin-wave excitation over the ground state. 
Clearly, the imaginary parts
of the roots $\lambda_j^{(1)}$ have conspired together and canceled the
$1/L^2$ correction proportional to the third mode. 
This is the first example we know that complex strings can produce 
negative contribution to the central charge. It is therefore tempting
to think the third excitation as a composite state of two elementary
spinons and that it does not contribute to the low-energy limit.
The same phenomenon happens to the excited states and this allows us
to interpret the operator content of the lowest excitation as
$X_{1}^{(1)}=X_1^{(2)}=1/4$. Putting these information together, we see
that the underlying
conformal field theory is likely to be represented in
terms of four Majorana fermions rather than given by
a $Sp(4)$ WZW theory. 

Let us now turn to the problem of extending our results for 
general $N>2$. The Bethe ansatz equations are parametrized by $N$ different
types of roots, and it turns out that the first $N-1$ roots are
given by both 1-strings and 2-strings while the last one behaves like
1-strings. If we characterize the center of the strings by $\xi_j^{(\alpha)}$,
$\alpha=1,\cdots, 2N-1$, we find that the corresponding density
distributions are
\begin{equation}
\sigma^{(\alpha)}(x)=
\left\{\begin{array}{l}
\frac{1}{N} \frac{\sin(\pi \alpha/N)}
{\cosh(2 \pi x/N) -\cos(\pi \alpha/N)},~\alpha=1, \cdots, N-1\\
\frac{1}{2(N+1)} \frac{1}
{\cosh[\pi x/(N+1)]},~ \alpha=N\\
\end{array}\right. 
\end{equation}
while  for
$\alpha=N+1,\cdots,2N-1$ they are given by the convolution
$\sigma^{(\alpha)}(x)=
\frac{1}{2\pi} \int_{-\infty}^{+\infty}
\sigma^{(2N-\alpha)}(y) \sigma^{(N)}(x-y) dy$.

The corresponding excitations are gapless, consisting of $N$
generalized spinons whose speed of sound are $v^{(\alpha)}=\frac{2 \pi}{N}$
for $\alpha=1, \cdots, N-1$ and $v^{(N)}=\frac{\pi}{N+1}$. In addition,
we have $N-1$ composite modes made by the convolution between the first
$N-1$ spinons with the $N$-th excitation. For $N>2$ numerical results
for large $L$ become difficult to be obtained since the number of roots
to be determined 
grow rapidly with both $N$ and $L$. However, for $N=3$ and small $L \sim 18$
our numerical analysis is consistent with the fact the only modes
contributing to the low-energy properties are the spinons, each one
with $c=1$. All these results seem to be strong evidence
that the continuum limit of such $Sp(2N)$ integrable models can indeed
be described in terms of $2N$ Majorana fermions.

In conclusion, we have studied the excitation spectrum of the simplest
integrable $Sp(2N)$ spin chain. Contrary to the common belief this system
is not the lattice realization of the $Sp(2N)$ WZW conformal theory.
Our study indicates that the nature of the excitations in spin-orbital
systems can be rather involving. In fact, the isotropic point 
$J_0=J_1=J_2$ is known to have three basic excitations \cite{SU},
being the lattice realization of 
a $SU(4)$ WZW field theory \cite{LW1}. However,
the anisotropic point $J_0/J_1=J_0/J_2=1/3$ has only two 
independent
excitations and one composite mode that do not contribute
to the low-energy limit, and it is described by a 
$c=2$
conformal field theory.
This work prompts us
to ask some questions that may open up new interesting avenues.
What is the nature of
the excitations of the spin-orbital model (1) in the crossover regime
$1/3 \leq J_0/J_1=J_0/J_2 \leq 1$? What is the mechanism that made
one of the excitations
to become a composite state? 
What is the integrable lattice $Sp(2N)$ model whose
continuum limit corresponds to the $Sp(2N)$ WZW theory?

\section*{Acknowledgements}
We thank A.L. Malvezzi
for valuable discussions and the hospitality of the Institute
of Advanced Study, Princeton, where part of this work has been carried out.
This work was supported by Brazilian agencies 
Fapesp and CNPq.

\newpage
\underline{Table 1}: Finite size sequences and the extrapolations 
of the effective central charge  
and the lowest conformal dimension of the $Sp(4)$ spin chain.

\begin{table}
\begin{center}
\begin{tabular}{|c|c|c|} \hline
     $L$  &$c_{ef}$  &$X_{ef}$    \\ \hline\hline
12 & 4.074 654 &1.000 985     \\ \hline
16 &4.046 720  &1.004 993     \\ \hline
20 &4.033 092  &1.006 011     \\ \hline
24 &4.025 305  &1.006 017     \\ \hline
28 &4.020 376  &1.005 692    \\ \hline
32 &4.017 024  &1.005 272   \\ \hline
36 &4.014 620  &1.004 845  \\ \hline
Extr. & 4.001 ($\pm 2$) & 1.006 ($\pm 2$)   
\\ \hline
\end{tabular}
\end{center}
\end{table}

\end{document}